\documentstyle[twocolumn,amssymb,aps,psfig,floats]{revtex}

\begin{document}
\draft

\twocolumn[\hsize\textwidth\columnwidth\hsize\csname @twocolumnfalse\endcsname

\title{Effects of long range electronic interactions on a 
one-dimensional electron system}
\author{Sylvain~Capponi and Didier~Poilblanc}
\address{Laboratoire de Physique Quantique \&
Unit\'e Mixte de Recherche 5626, C.N.R.S., \\
Universit\'e Paul Sabatier, 31062 Toulouse, France}
\author{Thierry~Giamarchi}
\address{Laboratoire de Physique des Solides \&
Unit\'e Mixte de Recherche 2, C.N.R.S.,\\
Universit\'e Paris-Sud, 91405 Orsay Cedex, France}

\maketitle

\begin{abstract}
\begin{center} 
\parbox{14cm}{
The effects of a long range electronic 
potential on a one dimensional chain of spinless fermions are investigated
by numerical techniques (Exact Diagonalisation of rings with up to
30 sites complemented by finite size analysis) and analytic calculations. 
Due to a competition between the $2 k_F$ oscillations in the density
and the (very slow) log divergence of the long wavelength part of the 
Coulomb potential, the metallic character of the system is enhanced at
intermediate (and up to quite large) lengthscales. 
Despite some similarities, we found that this quasi-metallic regime
is not of the Luttinger Liquid type, as evidenced by strong deviations from 
predictions of conformal field theory and in agreement with the picture of
a (very weakly pinned) Wigner crystal.
When the strength of the Coulomb interaction is substancially increased, we 
observe a smooth cross-over to a strongly localized charge density wave.
}
\end{center}
\end{abstract}

\vskip2pc]

\section{Introduction}

In one dimensional systems long range (LR) Coulomb potential 
plays a particularly important role in electronic models\cite{kondo}.
Indeed, as the dimensionality of an electronic system is reduced,
charge screening become less effective in
reducing the range of the electron-electron interactions. 
The importance of LR interactions between carriers has been 
highlighted in some experiments on 
$GaAs$ quantum wires\cite{goni} or quasi-one-dimensional
organic conductors\cite{Jerome_Schulz,dardel}. 
The role of a $1/r$ Coulomb repulsion on the long distance properties 
of electrons confined to a chain has been theoretically investigated using 
bosonisation techniques\cite{Schulz_crystal}. 
It was found that the charge correlations  decay very slowly with distance
suggesting that the ground state (GS) is similar to a classical 
Wigner crystal (WC) \cite{glazman_single_impurity,Schulz_crystal}. 
In this analysis
quantum fluctuations only lead to small oscillations of the particles 
away from the configuration of minimal potential energy.
However, this previous study was made in the continuum limit 
and the role of the lattice in this context is unclear.
Theoretical, both analytic\cite{giamarchi,giamarchi_mott_shortrev,%
Schulz_review1D,mori} or numerical\cite{Didier_Coulomb}, 
and experimental\cite{optical} results suggest that
umklapp processes have a crucial importance in 1D for the Mott transition.
In addition, the $q\rightarrow 0$ singular forward scattering 
interaction and the $2k_F$ backward scattering might lead to competing effects.
Although it is believed that, strictly speaking, for a long-range 
$1/r$ potential of arbitrary magnitude, the lattice would always produce 
a pinning of the WC, such an effect is (a) only achieved for very large 
systems (for not too large interaction) due
to its logarithmic nature and (b) is very weak. 
Hence, easy de-pinning of the WC can occur for {\it mesoscopic} systems
at small but finite temperature.  

In order to investigate the interplay between 
Coulomb interaction and umklapp scattering as well as their size dependent 
effects, thermodynamic observables are calculated
on chains with a variety of electron-electron interactions
using numerical techniques and compared to some analytic predictions
based on the bosonisation method. 
As the range of the electronic potential increases,
our computational analysis suggests that,
{\it at intermediate length scales}, the metallic character of the system
is enhanced. However, strong umklapp occurs at large Coulomb potential
leading to an insulating Charge Density Wave (CDW) state. 
Based on our numerical data and on analytic calculations we argue in favor of
a smooth cross-over between a weakly pinned WC (exhibiting metallic character
at length scales and energy scales we are dealing with here) and a
CDW insulator. 
We believe that the physical behaviors observed here for systems of
sizes $L\sim 20$-$30$ are of particular relevance for sub-nanometric 
quantum wires or carbon nanotubes.

This article is organised as follows: In Sec.~II we investigate the role
of an extended (but finite) range interaction. Spectacular oscillations
in e.g. the charge stiffness are observed numerically as the range of the
interaction is increased.  In Sec.~III, properties of a full $1/r$ long range
potential are analysed. First, the finite
size behaviors of the single particle and two particle charge gaps as well
as the charge stiffness are studied. 
Behaviors reminiscent of a Luttinger Liquid (with charge gaps $\propto 1/L$
and charge stiffness $\propto 1/L^2$\cite{LL_scaling}) are 
seen at small and intermediate
interactions. Qualitative changes occur at very large Coulomb repulsion with 
clear signatures of the opening of a charge gap and a drastic reduction of
the charge stiffness. Based on a conformal field theory analysis,
we then analyse the role of umklapp scattering by calculating the
(size dependent) anomalous exponent. In particular, we show that, even at
small Coulomb potential, strong deviations from the Luttinger Liquid
(LL) predictions occur.
Lastly, the momentum dependent low energy spectrum as well as
the single particle spectral function are computed confirming the
cross-over scenario.

The model analyzed here consists of a single periodic chain of spinless
fermions with $L$-sites and an extended interaction,
\begin{eqnarray} \label{ham}
H &=& -t\sum_{i} (c_{i}^\dagger c_{i+1} + h.c.) \\
&+& \sum_{i> j} V_{i-j} (n_i-{\bar n})(n_j-{\bar n}) \nonumber
\end{eqnarray}
where $n_i$ ($\bar n$) is the local (average) electron density, and the
rest of the notation is standard ($t=1$ is the unit of energy).
Two types of potentials will be considered:
(i) a plain LR Coulomb interaction
$V_{i-j}=V/|i-j|$;
(ii) an extended range potential, typically $V_{i-j}=V/|i-j|$ for 
$|i-j|\le r_{\mathrm max}$ 
and $V_{i-j}=0$ otherwise, extending up to a distance $r_{\mathrm max}$. 
Using Exact Diagonalization techniques, 1D closed chains are here analyzed.
These numerical results are compared to analytic predictions based on
the bosonisation technique.
A uniform ionic background has been included 
in Eq.~(\ref{ham}) such that the electrostatic energy per unit volume 
remains finite. 
In the rest of the paper, electron densities ${\bar n}=\frac{1}{2}$ and
${\bar n}=\frac{1}{3}$ will be considered. 
Finite size scaling analysis are performed by considering rings of 
typical sizes $L=12$, $16$, $20$, $24$ and $28$ [resp. $L=12$, $18$, $24$, 
and $30$] for ${\bar n}=\frac{1}{2}$ [resp. ${\bar n}=\frac{1}{3}$].
For an even number $N_e$ of
particles closed shell configurations are used by taking antiperiodic 
boundary conditions (ABC). 

\section{Finite range potentials}

\begin{figure}[htb]
\begin{center}
\mbox{\psfig{figure=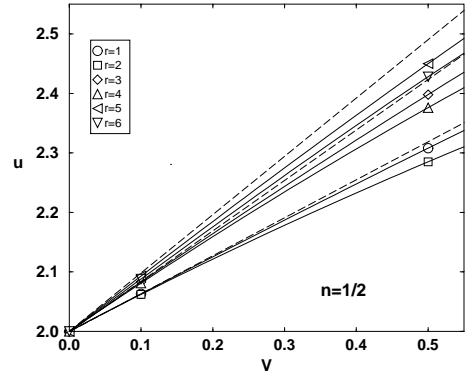,width=6cm,angle=-90}}
\end{center}
\caption{
Extrapolated velocity at density ${\bar n}=1/2$ vs $V$ for 
different ranges $r_{\mathrm max}\equiv r$ of the
electron-electron repulsion. The dotted lines correspond to the
theoretical predictions (\protect\ref{eq:kkeval}) for $r_{\mathrm max}=1$ or $2$,
$3$ or $4$, and $5$ or $6$.  
}
\label{velocity_ext}
\end{figure}

We start our analysis by first gradually varying the range of the 
electron interaction
$V_{i-j}$. For a (not too large) finite range interaction, the system is
expected to belong to the LL universallity class
characterized by a linear collective charge mode (see Appendix).
The corresponding charge velocity can easily be obtained from the slope of the 
low energy mode i.e. the energy difference between the first excited 
state at the smallest finite momentum $q=\frac{2\pi}{L}$ and the GS. 
Numerically, it is found that the expected $1/L^2$ behavior of the
finite size corrections \cite{LL_scaling}
is very well fulfilled so that a simple fit provides accurate 
extrapolated results shown in Fig.~\ref{velocity_ext} as a function of
the potential strength $V$. 
It is clear from this picture that the leading correction in $V$ is linear
and that the successive components $V_{2p}$ and $V_{2p+1}$ of the potential
have quite different effects on the increase of $u$ compared to
the non-interacting case ($u=v_F=2$ at half filling). 
Indeed, adding even distance interactions barely affect the slopes 
of $u$ vs $V$.

Some hint for this behavior can be found by using the continuum representation
explained in Appendix~\ref{appendix}. As shown in the Appendix, because of 
the presence of a $2k_F$ oscillations in the density, the various range 
components of the interaction interfere destructively. The 
bare Luttinger liquid parameters, describing the strength 
of the interactions are affected as (see (\ref{eq:kk}))
\begin{eqnarray} \label{eq:kktext}
u/K & = & v_F + \frac2{\pi}\sum_r V_r (1 - \cos(2k_Fr)) \\
uK & = & v_F \nonumber
\end{eqnarray}
For an half filled band $2k_F=\pi$ and $v_F=2$. 
In (\ref{eq:kktext}) one can thus distinguish
two contributions: (i) the one coming from the long wavelength part of the 
density ($q\sim 0$) for which all interactions simply add. This is the 
term that is responsible for divergence associated with the long 
range Coulomb interaction since $\sum_r V_r$ diverges logarithmically 
in this case; (ii) the contribution coming from the $2k_F$ part of the 
density. Because the density oscillates, interactions of different 
range can now interfere destructively. For long range interactions and 
large system sizes this term is bounded (it corresponds to the Fourier 
transform of the interaction potential with a Fourier component of $2k_F$)
and the system is thus totally dominated by the long wavelength part.
For finite size systems and/or finite range interactions the correction 
coming from the $2k_F$ term can be sizeable. In particular, at half filling
one would get 
\begin{eqnarray} \label{eq:kkeval}
u &\simeq& 2 + \frac1{\pi}(2V_1 + 2 V_3 + \cdots ) \\
K &\simeq& 1 - \frac1{2\pi}(2V_1 + 2 V_3 + \cdots) \nonumber
\end{eqnarray}
thus all even terms diseappear in $K$ and $u$ giving a $K$ [$u$] decreasing
[increasing] less rapidly than if only the $q\sim0$ part was 
concerned. In particular 
one would have $2 V$ in (\ref{eq:kkeval}) for a nearest neighbor 
interaction and only $\sim 4.36 V$ for a $22$ sites interaction. 

\begin{figure}[htb]
\begin{center}
\mbox{\psfig{figure=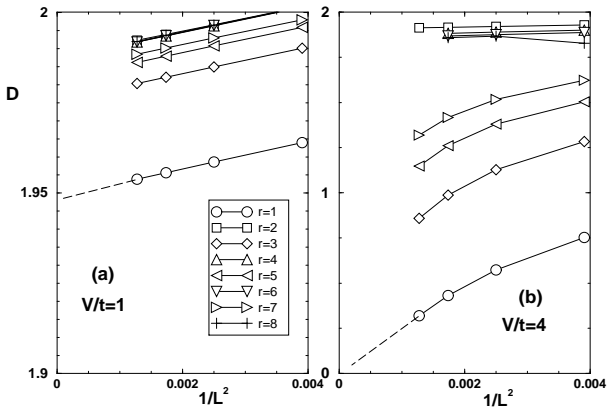,width=8cm,angle=-90}}
\end{center}
\caption{
Charge stiffness at density ${\bar n}=1/2$ vs inverse square length for 
different ranges $r_{\mathrm max}\equiv r$ of the
electron-electron repulsion with amplitudes (a) $V=1$ and (b) $V=4$.
}
\label{drude_ext}
\end{figure}

The charge stiffness $D$ can be obtained numerically 
by threading a flux $\Phi$ (in unit of the
flux quantum) through the ring. The curvature around the minimum of the
GS energy $E_0(\Phi)$ provides the best estimate, namely 
$D=\frac{1}{2}\partial^2(E_0/L)/\partial\phi^2)$ 
[where $\phi$ is a pseudo momentum 
$\phi=(2\pi/L)\Phi$]. Note that, for an even number of electrons, the
minimum occurs at $\Phi=1/2$ i.e. for antiperiodic boundary conditions  
corresponding to a non-interacting closed shell configuration. 
Results obtained on $L=16$, $20$, $24$ and $28$-site rings with 
$N_e=L/2$ electrons are shown in Fig.~\ref{drude_ext}.
We observe that the charge stiffness for $r_{\mathrm max}>1$ 
is systematically {\it enhanced} compared to the case of
nearest neighbor (NN) repulsion ($r_{\mathrm max}=1$). Note that 
for $r_{\mathrm max}=1$ a metal-insulator transition is known 
exactly to occur at $V=2$ so that $D$ in Fig.~\ref{drude_ext}(a) 
[resp. Fig.~\ref{drude_ext}(b)] extrapolates to a finite [resp. vanishing]
value in that case. An interesting oscillatory behavior appears as
$r_{\mathrm max}$ increases. For $r_{\mathrm max}=2p+1$ (odd),
$D$ monotonically {\it increases} with $p$. The maximum of the charge stiffness
is obtained for $r_{\mathrm max}=2$ and $D$ {\it decreases} slightly
for larger $r_{\mathrm max}=2p^\prime$ (even). The (unique) limit corresponding
to a long range $1/r$ 
repulsion would then be reached from below (above) when $p\rightarrow\infty$ 
($p^\prime\rightarrow\infty$). We believe that this feature is generic and 
is independent of the magnitude of $V$ as long as the potential is 
{\it short range} i.e. $r_{\mathrm max}\ll L$. 

Our numerical results for $D$ can also be explained qualitatively
using the bosonisation picture. Indeed, when more components $V_p$ of the
potential are added, in addition to the change of the LL 
parameter $K$ described by (\ref{eq:kk}), a dramatic effect occurs 
also in the strength of the umklapp term (see Appendix) 
\begin{equation}
g = \sum_r V_r \cos(2k_F r)
\end{equation}
At half filling, this term is {\it reduced} for all 
even range contributions (since one has 
$V_1-V_2$, $V_1-V_2+V_3-V_4$, etc. thus enhancing 
the metallic character of the system for intermediate 
sizes)\cite{note1}. Since without such umklapp term the stiffness $D$
is not renormalized by interactions $D=uK=v_F$ (see (\ref{eq:kktext})), 
the strength of the umklapp is a direct measure of the reduction of the 
stiffness due to interaction effects (the stronger the umklapp term, 
the smaller the stiffness). 
For a second neighbor interaction the umklapp 
strength $g$ is at its minimum $V/2$, consistent with the numerical 
findings of the largest stiffness. Note also that for second neighbor 
interactions $K$ is not decreased compared to the nearest neighbor one, 
thus allowing for a maximum enhancement of the charge stiffness. 
For odd range interactions the serie 
$V_1$, $V_1-V_2+V_3$, $V_1-V_2+V_3-V_4+V_5$ and so on is monotonically
decreasing, whereas for even ranges $V_1-V_2$, $V_1-V_2 
+V_3-V_4$ and so on is increasing corresponding respectively to increasing 
and decreasing charge stiffness in good agreement with the numerical results.
When the whole serie is included (for the Coulomb case) the umklapp strength
is $g \sim 0.69 V$ and thus much weaker than the nearest neighbor one 
$g=V$. 
 
Of course for larger length scales (\ref{eq:kktext})
is dominated by the logarithmic divergence coming from the $q\sim 0$ component
whereas the $2k_F$ part gives a finite contribution and $K\to 0$. Because 
of the presence of the umklapp term this always lead to an insulating gapped 
phase ($K< K_c$). However both for intermediate sizes or finite range 
interactions the presence of long range interactions is in fact favorable 
to the metallic behavior.

\section{Coulomb 1/\lowercase{r} potential}

\subsection{Charge gaps}

\begin{figure}[htb]
\begin{center}
\mbox{\psfig{figure=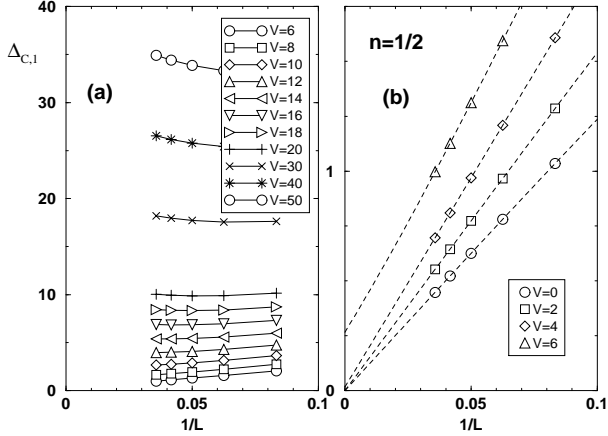,width=8cm,angle=-90}}
\end{center}
\caption{
Single particle charge gap vs $1/L$ for ${\bar n}=1/2$ and several 
values of the LR Coulomb interaction $V$. Dotted lines are guides
to the eye. 
}
\label{1Pgap}
\end{figure}

\begin{figure}[htb]
\begin{center}
\mbox{\psfig{figure=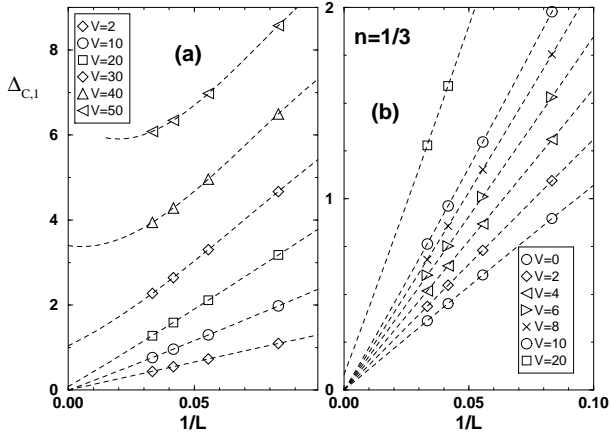,width=8cm,angle=-90}}
\end{center}
\caption{
Single particle charge gap vs $1/L$ for ${\bar n}=1/3$ and several 
values of the LR Coulomb interaction $V$. Dotted lines are guides
to the eye. 
}
\label{1Pgap_n033}
\end{figure}

We now turn to the case of the full $1/r$ potential and to the investigation
of the metal-insulator transition.
The single particle charge gap at average density ${\bar n}=\frac{N_e}{L}$ 
is defined by $\Delta_{C,1}=E_0(N_e+1)+E_0(N_e-1)-2E_0(N_e)$.
Note that the GS energies $E_0(N_e\pm 1)$ are calculated 
by keeping the ionic charge density constant (i.e. ${\bar n}$ fixed in 
Hamiltonian~(\ref{ham})). 
Our results for $\Delta_{C,1}$ at ${\bar n}=1/2$ [resp. ${\bar n}=1/3$] 
are shown on Fig.~\ref{1Pgap} [resp. Fig.~\ref{1Pgap_n033}].

At small and intermediate interactions, the charge gap seems to 
vanish as $\sim 1/L$, a behavior characteristic of a Luttinger Liquid (LL).
However, we cannot exclude a very small extrapolated value. 
This suggests that the GS has a metallic character, at least approximately
(i.e. with a very small charge gap). On the other hand, 
at large Coulomb repulsion, a charge gap clearly opens up. 
Note that at lower electron commensurability than half 
filling, a significantly larger interaction is needed to clearly see 
a signature of the gap: $V\sim 20$ at ${\bar n}=1/3$ compared to
only $V\sim 5$ at ${\bar n}=1/2$.

\begin{figure}[htb]
\begin{center}
\mbox{\psfig{figure=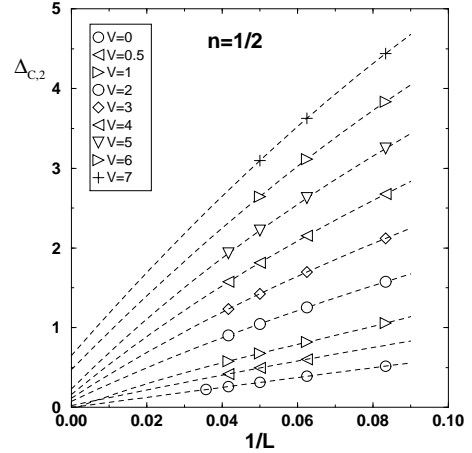,width=6cm,angle=-90}}
\end{center}
\caption{
Two-particle charge gap vs $1/L$ for ${\bar n}=1/2$ and several 
values of the LR Coulomb interaction $V$. Dotted lines are guides to the eye. 
}
\label{2Pgap_vsL}
\end{figure}

In order to estimate the sensitivity of the results to shell effects,
we have also computed the {\it two-particle} charge gap at 
${\bar n}=1/2$ defined by $\Delta_{C,2}=E_0(N_e+2)+E_0(N_e-2)-2E_0(N_e)$ 
which only involves {\it even} number $N_e$ of particles.
Results are shown in 
Fig.~\ref{2Pgap_vsL}. For this range of parameters and system sizes,
we observe, as expected, that $\Delta_{C,2}$ approximatively 
equals $2\Delta_{C,1}$ although the additional finite size correction 
(typically $\sim 1/L^2$) on top of the leading $1/L$ term have a different 
sign.

\begin{figure}[htb]
\begin{center}
\mbox{\psfig{figure=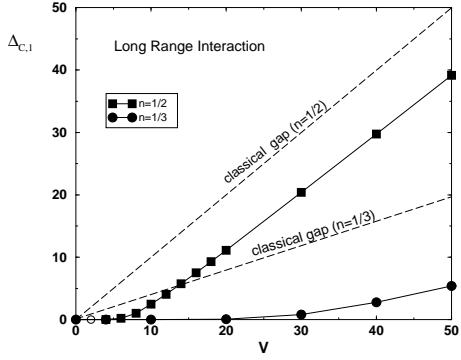,width=6cm,angle=-90}}
\end{center}
\caption{
Single particle charge gap for ${\bar n}=1/2$ and ${\bar n}=1/3$
as a function of the magnitude of the LR Coulomb interaction $V$. 
The dotted lines correspond to the classical behaviors
$\Delta_{C,1}\propto V$ valid in the $t=0$ limit. 
}
\label{1PgapvsV}
\end{figure}

Extrapolations of our data for $\Delta_{C,1}$ to the thermodynamic limit 
have been attempted
using a cubic spline (including higher powers of $1/L$ if necessary) and the 
results are shown in Fig.~\ref{1PgapvsV}. In the classical limit i.e. $t=0$,
electron (hole) excitations are multiple massive fractional charge excitations 
(solitons). For example, at half filling a dephasing of $\pi$ in the 
01010101 etc. CDW pattern like 01011010 etc. produces a charge-1/2 excitation.
An extra electron introduced in the system will then split into two such 
elementary excitations.
Our numerical results at large $V$ agree well with the classical estimates e.g.
$\Delta_{C,1}=V$ at ${\bar n}=1/2$ or 
$\Delta_{C,1}=(1-\frac{\pi\sqrt{3}}{9})V$ at ${\bar n}=1/3$\cite{Hatsugai}.
We observe that a sizeable gap opens up in the LR model only at 
values of $V$ significantly larger than the critical values
corresponding to the metal-insulator transitions of the related ``screened'' 
models where the range of the interaction is limited to 
$r_{\mathrm max}=\frac{1}{\bar n}-1$ ($\frac{1}{\bar n}$ being the average
inter-particle distance). For example, at ${\bar n}=1/2$, the single particle
gap for a LR potential of magnitude $V$ ($V>4$) is roughly of the same 
order than the gap produced by a short range NN interaction of 
magnitude $V/2$. This clearly shows again that the effect of umklapp 
scattering is reduced in the presence of a LR interaction. 
However, we believe that, in the case of a LR $1/r$ potential,
a cross-over rather than a real transition occurs for increasing $V$.

\subsection{Charge stiffness}

\begin{figure}[htb]
\begin{center}
\mbox{\psfig{figure=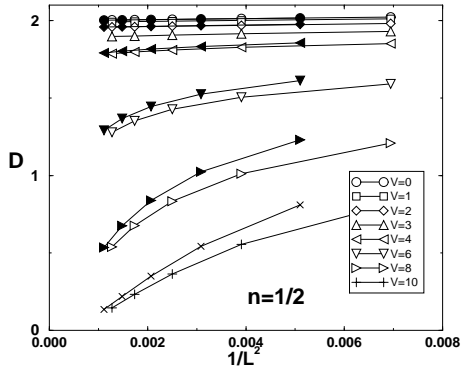,width=6cm,angle=-90}}
\end{center}
\caption{
Charge stiffness at density ${\bar n}=1/2$ vs inverse square length for 
a $1/r$ long range potential of several magnitudes (as indicated on the
picture). 
Open (closed) symbols correspond to an even (odd) number 
of electrons.
}
\label{drude_lr}
\end{figure}

Results for the charge stiffness of the $1/r$ model at density $n=1/2$
are shown in Fig.~\ref{drude_lr}. Note that we have used here two sets 
of closed rings of sizes $L=4p$ ($12$ to $28$ sites) and $L=4p+2$
($14$ to $30$ sites). ABC and PBC are used
respectively to ensure ($V=0$) closed shell configurations in all cases. 
However, we observe that the two sets of data exhibit slightly 
different scaling behaviors which can be attributed to
small shell effects. In agreement with the previous results for the
single particle gap, these data suggest an effective metallic behavior at
small and intermediate magnitude $V$ while for fairly large $V$ the
charge stiffness is suppressed, even at these length scales.
Note that, for, let us say, $V\le 4$ the charge stiffness is
particularly large, typically of the same order as in a 
non-interacting system. Since $D$ is directly proportional to the Drude weight
in the optical conductivity, this signals that most of the optical weight
lies at $\omega\sim 0$ (eventhough a tiny pinning frequency might exist).
This contrasts with the large transfer of weight to finite frequencies 
occuring in the insulating phase of the NN t-V model for $V>2$. 

\subsection{Anomalous exponents}

Although the singular $q\rightarrow 0$ part of the $1/r$ potential
does not alone lead to true localisation, it is nevertheless 
responsible to a $K\rightarrow 0$ LL constant so that umklapp scattering
becomes relevant and can lead to an insulating behavior. 
However, from the previous analysis, this effect is
quite small at intermediate coupling $V$ and for the system
sizes we considered here. Hence, we shall attempt here a more sensitive
analysis based on identities from conformal field theories.
Note that the latter are {\it a priori} valid only in the case of 
standard Luttinger liquids (i.e. 1D models with short range interactions).
Therefore, deviations from these expected relations will 
signal non LL behaviors. 

\begin{figure}[htb]
\begin{center}
\mbox{\psfig{figure=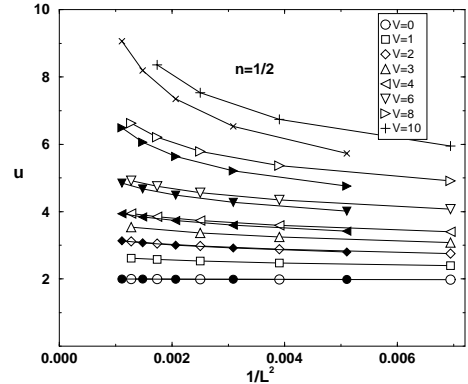,width=6cm,angle=-90}}
\end{center}
\caption{
Charge velocity $u$ in the long range $1/r$ model at density ${\bar n}=1/2$
plotted as a function of the inverse square length. 
Open (closed) symbols correspond to an even (odd) number 
of electrons.
}
\label{velocity_lr}
\end{figure}

To perform this analysis the charge velocity is needed.
Results are shown in Fig.~\ref{velocity_lr}. The charge velocity increases 
with $V$ and system size as expected. In the continuum limit,
the WC picture predicts a low energy dispersion of the form 
$q\log^{1/2}{(q)}$. This logarithmic divergence comes directly 
from the $q\sim 0$ contribution of the potential to $u/K$ as shown in 
(\ref{eq:kktext}) since $\sum_r V_r$ is logarithmically divergent. We 
thus expect a very weak 
$\log^{1/2}{(1/L)}$ divergence of $u$ consistent with our data
at sufficiently large $V$. 

The anomalous exponent $K$ can be calculated for each ring size $L$
by using the conformal identity $K=\frac{D}{u}$. Such an identity 
can be easily derived from the quadratic form (\ref{eq:quadratic}) of the 
boson representation. This quantity is of particular importance since, 
in a conformal 
invariant theory, the charge correlations decay as a power law whose exponent
is precisely given by $K$. It will also signal, as shown in the appendix,
when an umklapp term becomes relevant and leads to a gap in the spectrum
and an insulating behavior. 

\begin{figure}[htb]
\begin{center}
\mbox{\psfig{figure=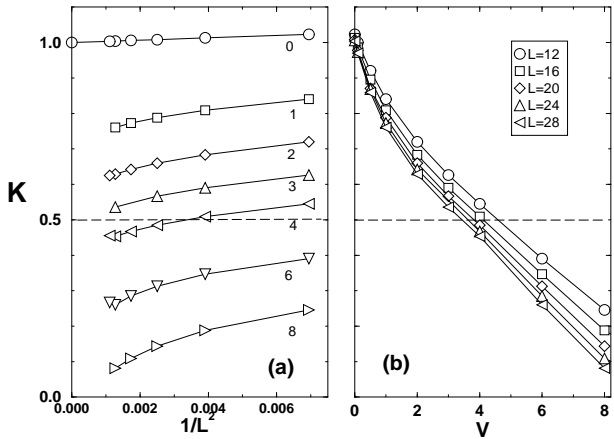,width=8cm,angle=-90}}
\end{center}
\caption{
Anomalous exponent $K$ (calculated from $D$ and $u$) 
of the long range $1/r$ model at density ${\bar n}=\frac{1}{2}$. 
(a) Finite size behavior for different magnitudes $V$
(as shown on the plot). Note that the data 
obtained on the largest $L=30$ ring scale slightly differently from the
data set obtained on $L=4p$ rings; 
(b) vs $V$ for different sizes (as indicated
in the legend). The critical value $K_c=1/2$ at half filling is 
shown by dotted lines on each plot. 
}
\label{K_lr}
\end{figure}

Our data for $K$ at ${\bar n}=1/2$ obtained for various Coulomb repulsion $V$ 
are plotted vs $1/L^2$ in Fig.~\ref{K_lr}(a). The same set of data is shown in 
Fig.~\ref{K_lr}(b) as a function of $V$. For large $V$,
$K<K_c$ so that umklapp scattering is expected to be strong in this regime 
and to produce a large charge gap in agreement with the previous findings.
The results at small and intermediate $V$ are not conclusive but suggest
that umklapp is weak in this regime. However, the (slow) 
divergence of the charge velocity should ultimately lead to
$K(L)\rightarrow 0$ and then, eventually, to the opening of a very small
charge gap at large sizes. Note that 
$K$ seems to be a more sensitive probe than the single particle gap itself.
Indeed, the data in Fig.~\ref{K_lr}(a) for e.g. $V=4$ unambiguously
show the relevance of umklapp while the extrapolation of 
$\Delta_{C,1}$ in Fig.~\ref{1Pgap}(b) for the same parameter is still
unconclusive. 

\begin{figure}[htb]
\begin{center}
\mbox{\psfig{figure=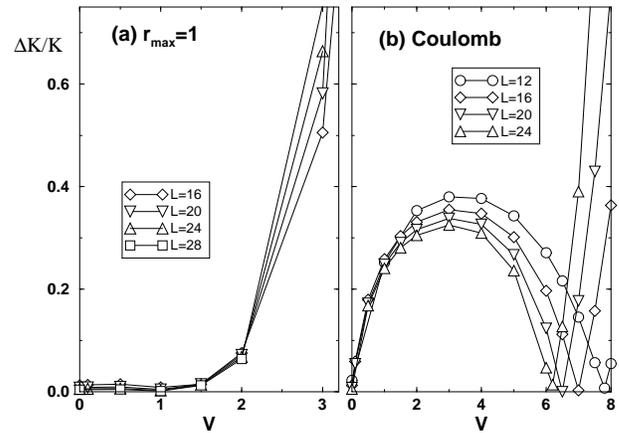,width=8cm,angle=-90}}
\end{center}
\caption{
Relative error $\Delta K/K$ versus $V$ at density 
${\bar n}=\frac{1}{2}$ for a short range NN repulsion (a)
and a long range Coulomb interaction (b). 
}
\label{DeltaK_lr}
\end{figure}

One can get additional insights on the possible instability of the Luttinger
liquid fixed point by considering a different 
estimation $K'(L)$  (see e.g. \cite{estimation_K}) of the anomalous exponent
$K^\prime=\pi u \kappa^{-1}$, where $\kappa^{-1}$ is the inverse 
compressibility which can be estimated numerically as 
$\kappa^{-1}(L)=\frac{L}{4}\Delta_{C,2}$ in order to avoid shell effects.
In a regular LL the two estimates $K$ and $K^\prime$ converge toward the
same thermodynamic limit within finite size corrections $\sim 1/L^2$. 
In order to examine the breakdown of the LL character, it is then of interest
to define the relative difference, 
\begin{equation}
\Delta K/K=|1-\frac{\pi u^2}{\kappa D}|\, . 
\end{equation}
This quantity is plotted vs $V$ in Figs.~\ref{DeltaK_lr}(a) and (b)
for the NN and the long range model respectively using the same scale
for a direct comparison. As expected, in the NN model for $V<2$ the
relative ratio $\Delta K/K$ is very small (typically below $0.01$)
and strongly increases for $V>2$ in the insulating CDW phase.
The situation is drastically different in the LR model where 
$\Delta K/K$ is much larger for all $V$. Note that the minimum around
$V\simeq 6$ corresponds in fact to an accidental change of sign of 
the difference $K-K^\prime$ and moves continuously towards small 
$V$ for increasing size. 
We believe that Fig.~\ref{DeltaK_lr}(b) provides strong evidence
that the quasi-metallic phase of the LR model is not
a simple LL fixed point. It would be consistent with a system 
which would be a pinned (by the lattice) WC in the infinite size
limit and which remains metallic of the finite size effects  
\cite{Didier_Coulomb}. 

\subsection{Low energy modes and dynamics}

\begin{figure}[htb]
\begin{center}
\mbox{\psfig{figure=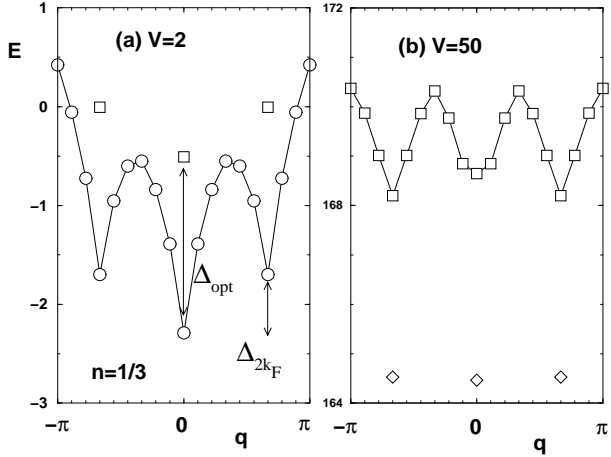,width=8cm,angle=-90}}
\end{center}
\caption{
Low energy spectrum of the LR Coulomb model at ${\bar n}=1/3$ vs momentum $q$.
At momenta $0$ and $2\pi/3$ the first excited state is also shown.
The definitions of the finite size gaps 
$\Delta_{\mathrm opt}$ and $\Delta_{2k_F}$
are shown on the plot. (a) $V=2$; (b) $V=50$. 
}
\label{spectrum_lr}
\end{figure}

Analysing the structure of the low energy spectrum for all $q$ is useful
to get further understanding on the quasi-metal -- insulator cross-over.
The dispersion of the GS vs momentum is shown in Figs.~\ref{spectrum_lr}
at constant particle density ${\bar n}=1/m=1/3$ and for two extreme values of 
the Coulomb repulsion, $V=2$ and $V=50$. Although one observes clearly 3 
local minima at momenta $0$ and $\pm 2\pi/3$ in both cases, the two spectra  
are qualitatively very different. While at small $V$ linear branches
seem to appear in the vicinity of the minima, the three almost degenerate
GS are clearly separated from the rest of the spectrum at large 
$V$\cite{note2}. This last feature is characteristic of translation 
symmetry breaking occuring in a $2k_F$ CDW state in which the 
(spinless) particles are localised at equal spacings.

\begin{figure}[htb]
\begin{center}
\mbox{\psfig{figure=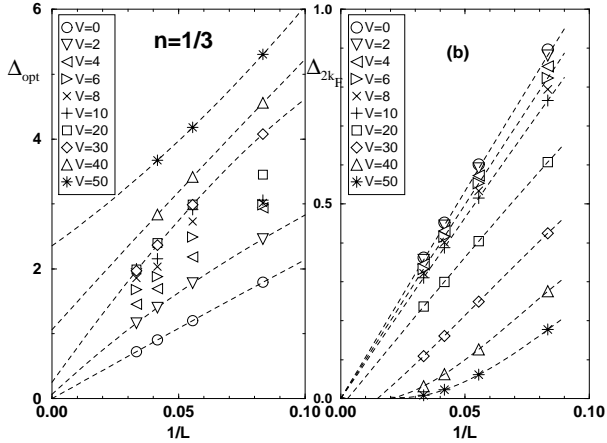,width=8cm,angle=-90}}
\end{center}
\caption{
Finite size scaling behaviors of the optical (a) and $2k_F$ (b) gaps
vs $1/L$ for ${\bar n}=1/3$.
}
\label{gaps_lr}
\end{figure}

A more qualitative analysis can be performed by studying the size dependence of
the gap between the GS and the next excited state at momentum $q=0$
(so-called optical gap) which is shown in Fig.~\ref{gaps_lr}(a).
At small $V$, $\Delta_{\mathrm opt}$ extrapolates for $L\rightarrow\infty$ 
to a vanishing or small value characteristic of a metallic or quasi-metallic
behavior. On the contrary, the extrapolated gap is clearly finite 
for large $V$ as expected in the case of an insulator. 
Note that, for intermediate $V$, we observe some irregular behavior of the data
so that no finite size scaling analysis was attempted here.
Let us point out again that the absence of gap for small $V$ is related 
to the finite size of the system and the fact that neither the velocity 
$u$ nor the Luttinger parameter $K$ have been renormalized to zero because 
of the log divergence of $\sum_r V_r$. 

The instability towards a translation symmetry breaking CDW state 
can be also evidenced from the energy separation $\Delta_{2k_F}$
between the absolute GS and the lowest energy at momentum $2k_F$ (see 
Fig.~\ref{spectrum_lr}(a) for the definition). As seen in 
Fig.~\ref{gaps_lr}(b), at small $V$ $\Delta_{2k_F}$ scales like
$1/L$, a behavior reminiscent of a LL. Expected deviations from this behavior 
due to logarithmic corrections can not be seen for such system sizes.
For large $V$, $\Delta_{2k_F}$ decreases faster than $1/L$. 
The data for $V=40$ and $V=50$ are compatible with the expected exponential 
behavior for system sizes larger than the finite correlation length of 
the insulating CDW. 

\begin{figure}[htb]
\begin{center}
\mbox{\psfig{figure=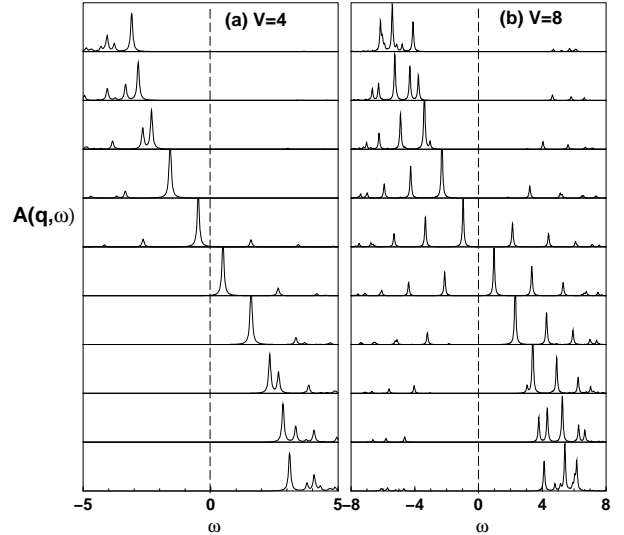,width=8cm,angle=-90}}
\end{center}
\caption{
Single particule spectral function vs frequency calculated on a $L=20$ sites
Coulomb ring at half filling. From top to bottom, the momenta increase
by equal steps from $0$ to $\pi$. Energies are measured from the chemical 
potential. (a) $V=2$; (b) $V=8$. 
}
\label{Akw_lr}
\end{figure}

Lastly, we have computed the single particle spectral function
$A(q,\omega)=-\frac{1}{\pi}{\mathrm Im} G(q,\omega)$, where $G(q,\omega)$
is the standard two-points Green function, and results at ${\bar n}=1/2$
for $V=4$ and $V=8$ are shown in Figs.~\ref{Akw_lr}(a) and (b) respectively. 
For $V=4$ most of the spectral weight is located in a single dispersive 
branch. However, for $V=8$ when the CDW correlations become stronger,
one observes clearly a quasi-symmetric ``image'' branch due to Bragg 
reflection on the CDW (fluctuating) potential. These data give a further 
confirmation of the cross-over occuring as the magnitude of the
LR Coulomb potential increases. 

\section{Conclusions}

To conclude, we have shown that, for intermediate lengthscales an 
interesting phenomenon occurs for a one dimensional system in the 
presence of long range interactions. In the infinite size 
limit, such a system would always be insulating due to the 
log divergence of the long wavelength part of the Coulomb potential
$\sum_r V_r$. However, because this divergence is quite slow, for 
intermediate (and up to quite large !) lengthscales an opposite effect 
occurs due to the $2 k_F$ oscillations in the density: long range 
interactions {\it enhance} the metallic character of the system.
Quite generically, umklapp scattering
is reduced (but not totally suppressed) by Coulomb LR interactions. 
At not too large $V$ (typically $V<20$ at $n=1/3$) the 1D electron
system exhibits quasi-metallic behaviors with vanishing
(or tiny) mass gaps in agreement with previous findings on spinfull LR 
models\cite{Didier_Coulomb}.
This phase, however, shows deviations from conformal field theory
predictions. A physical description in term of weakly pinned WC 
is proposed. At larger Coulomb interaction a cross-over towards 
a strongly localised $2k_F$ CDW occurs. Analytical calculations
based on bosonisation techniques confirm our numerical findings.

We thank IDRIS (Orsay) 
for allocation of CPU time on the C94 and C98 CRAY supercomputers.
TG would like to thank the ITP where part of this work was completed 
for support (under NSF grant No. PHY94-07194) and hospitality.

Note added: While completing this work we became aware of related work
by G.~Fano et al. [cond-mat/9909140] in agreement with our
analysis in the small $V$ regime.

\appendix
\section{Bosonisation}
\label{appendix}

One can study the effect of Coulomb interations on the 1d electron gas by 
using a boson representation of the fermion operators (bosonisation)
\cite{haldane_bosonisation,schulz_houches_revue}.
Introducing a field $\phi$ to represent the long wavelength fluctuations 
of the density, the kinetic energy in (\ref{ham}) becomes 
\begin{equation} \label{eq:quadratic}
H = \frac{1}{2\pi}\int dx u K (\pi \Pi)^2 + \frac{u}{K}(\nabla\phi)^2
\end{equation}
where $\phi$ and $\Pi$ are canonically conjugate operators. For the free
($V=0$)
Hamiltonian $u=v_F$ and $K=1$. The parameters $u$ and $K$ are the only one 
necessary to describe the low energy properties of the metallic phase 
and are known as Luttinger liquid parameters. They can be computed in 
perturbation in the interaction $V$ by using the expression of the density 
\begin{equation} \label{eq:density}
\rho(x) = \rho_0 -\frac1\pi\nabla\phi + \frac{1}{2\pi\alpha}e^{i(2k_Fx+2\phi)}
\end{equation}
where $\alpha$ is a short distance cutoff that can be identified with the 
lattice spacing. 
Using (\ref{eq:density}) for the interaction term in (\ref{ham}) one gets 
\begin{eqnarray} \label{eq:hint}
H_{\rm int} &=& \sum_{i,j} V_{i-j} \{ 
\frac1{\pi^2} (\nabla\phi_i)(\nabla\phi_j)\\
&+& \frac{1}{(2\pi\alpha)^2} 2 \cos(2k_F (x_i-x_j) + 2\phi_i-2\phi_j) \} \, .
\nonumber
\end{eqnarray}
By considering that the fields $\phi$ are slowly varying at the scale of the 
lattice one can expand (\ref{eq:hint}). To expand safely the fields should be 
normal ordered using 
\begin{equation} \label{eq:normal}
e^{i2(\phi(r_1)-\phi(r_2))} = :e^{i2(\phi(r_1)-\phi(r_2)}:
e^{-\frac12\langle[2(\phi(r_1)-\phi(r_2))]^2\rangle} \, .
\end{equation}
At the lowest order in $V$, (\ref{eq:normal}) becomes
\begin{equation} 
e^{i2(\phi(r_1)-\phi(r_2))} = :e^{i2(\phi(r_1)-\phi(r_2)}:
\frac{\alpha^2}{(r_1-r_2)^2}
\end{equation}
In the normal product one can expand $\phi(r_1)-\phi(r_2) \simeq
(r_1-r_2)\nabla\phi(R)$ where $R=(r_1+r_2)/2$. One thus 
recovers a quadratic form similar 
to (\ref{eq:quadratic}) but with 
\begin{eqnarray}\label{eq:kk}
u/K & = & v_F + \frac2{\pi}\sum_r V_r (1 - \cos(2k_Fr)) \\
uK & = & v_F \nonumber
\end{eqnarray}
where $r=1,2,3,...$ is the distance on the lattice. The above 
expression is valid at the lowest order in $V$. For finite $V$
the various coefficients in it will get renormalized by irrelevant 
operators. For a long range potential the $q\to0$ part is dominant and 
gives rise to the logarithmic divergence of $A_q$ responsible for the 
one dimensional plasmon. Due to the $2 k_F$ oscillations in density
the $2k_F$ component of the potential also occurs. Note that although 
(\ref{eq:kk}) looks superficially similar to what one could obtain 
using g-ology arguments, such arguments would naively suggest 
that $g_1=-g_2$. This would lead the $V(2k_F)$ part to enter both 
the $u/K$ and $uK$ term and lead to an incorrect renormalization of 
the $u K$ term. Such a change of $u K$ is impossible for a well 
defined lattive model \cite{giamarchi_umklapp_1d}, where one 
should have $uK=v_F$ as a  consequence of Galilean invariance in the 
absence of an underlying lattice. 

In the presence of a lattice one 
should take into account that the momentum is only conserved modulo
a vector of the reciprocal lattice (i.e. for example for half filling 
$4k_F=2\pi$). When taking this effect into account (\ref{eq:hint}) gives 
also rise to the so-called umklapp terms. These terms are the one 
responsible for the Mott transition and the opening of the charge 
gap \cite{giamarchi_mott_shortrev}. 
They can also be computed, and read e.g. for the half filled case
\begin{equation} \label{eq:umklapp}
H_{\rm umk} = \frac{-2 g}{(2\pi\alpha)^2} \int dx \cos(4\phi(x))
\end{equation} 
where 
\begin{equation}
g = \sum_r V_r \cos(2k_F r) \, .
\end{equation}
Taken together with (\ref{eq:quadratic}), (\ref{eq:umklapp}) is able to 
open a gap in the spectrum and lead to a Mott insulator (or equivalently 
to a Wigner crystal pinned on the lattice) for $K<K_c$, where $K_c$ depends 
on the commensurability. For half filling $K_c = 1/2$ and would be 
$K_c = 1/(2n^2)$ for a commensurability of order $n$.


\begin{references}
\bibitem{note1} As seen in Ref.~\protect\cite{Didier_Coulomb}, if only $V_1=V$ 
and $V_2=V/2$ exist, the metallic LL phase remains stable up to $V\simeq 5.5$.
\bibitem{note2} Similar behaviors have been seen for other LR models
at densities ${\bar n}=1/m$; see Ref.~\protect\cite{Hatsugai}.
\end{references}
\end{document}